\documentclass[
twocolumn,
]{ceurart}

\usepackage{listings}
\usepackage{graphicx} % For including graphics
\usepackage{url} % For handling URLs in captions
\usepackage{subfig} % For subfigures
\usepackage{algpseudocode}
\usepackage[linesnumbered,ruled,vlined]{algorithm2e} 
\usepackage{enumitem}
\usepackage{booktabs}
\usepackage{amsmath}
\SetKwInput{KwIn}{Input}      % defines \KwIn
\SetKwInput{KwOut}{Output}    % defines \KwOut
\definecolor{ImprovedRow}{HTML}{E8F5E9}

\begin{document}

\copyrightyear{2025}
\copyrightclause{Copyright for this paper by its authors.
  Use permitted under Creative Commons License Attribution 4.0 International (CC BY 4.0).}
\conference{}

\title{Enhancing LLMs through human feedback: a journey towards self-improvement}

\author[1]{Tatiana Pelc}
\author[1]{Gila Kamhi}
\author[1]{Asaf Avrahamy}
\author[1]{Adi Fledel-Alon}[ %
%orcid = 0009 - 0008 - 6305 - 9786 ,
email =adi.fledel.alon@intel.com ,
]
\cormark[1]
\address[1]{Intel Corporation}

\cortext[1]{Corresponding author.}

\begin{abstract}
In the rapidly evolving landscape of information retrieval systems, the ability to adapt and improve through user feedback is paramount. This study introduces a novel methodology for refining the performance of a \textit{primary} Retrieval Augmented Generation (RAG) system by strategically integrating an \textit{auxiliary} feedback RAG system. By systematically harnessing human-generated feedback, the approach aims to enhance the accuracy, relevance, and overall quality of responses, driving the system towards self-improvement. Central to this methodology is a human-in-the-loop implementation, where user feedback is continuously collected, classified, and integrated into the inference workflow, enabling the system to learn and evolve iteratively. To validate the effectiveness of this approach, the study employs rigorous testing against three diverse benchmark datasets focused on general and custom domain knowledge, utilizing a \textit{LLM-as-a-Judge} evaluation strategy. This comprehensive framework not only underscores the transformative potential of feedback-driven enhancements in RAG systems but also sets a precedent for future research in adaptive information retrieval technologies, marking a significant step in the journey towards autonomous refinement and optimization through user engagement.
\end{abstract}


\maketitle

\section{Introduction}

Large Language Models (LLMs) have driven recent advances in AI, enabled by their unprecedented scale and improved through Reinforcement Learning from Human Feedback (RLHF). RLHF refines model behavior using human input, significantly enhancing adaptability, responsiveness, and overall performance.

However, traditional AI systems have relied on binary feedback (e.g., like/dislike), which limits learning depth. This paper proposes incorporating rich textual human feedback—including detailed corrections—into the generative AI pipeline. This enables more nuanced learning and improves response accuracy and contextual alignment.

\subsection{Related work}\label{sec:relatedwork}

Retrieval Augmented Generation (RAG) enhances LLMs by retrieving relevant documents, but often fails to capture user preferences, resolve contextual mismatches, or adapt to evolving information needs. Common issues include retrieval bias, robustness challenges, and fairness and privacy concerns \cite{barnett2024sevenfailurepointsengineering, ni2025trustworthyretrievalaugmentedgeneration}. Early approaches with static retrieval and binary feedback offered limited gains.
Recent work emphasizes the role of structured and dynamic feedback to improve alignment and reliability. Hybrid methods combining human judgment with automated evaluation have been explored to reduce hallucinations. Human-in-the-loop systems and cross-domain adaptation are emerging as effective strategies for building self-improving, context-aware RAG systems.

\subsubsection{Feedback in RAG and LLM systems}

Structured feedback pipelines such as CDF-RAG (Causal Dynamic Feedback for Adaptive Retrieval-Augmented Generation) refine retrieval and generation through iterative loops guided by causal graphs, enabling multi-hop reasoning and improving accuracy \cite{khatibi2025cdfragcausaldynamicfeedback}.  In educational applications, timestamp-anchored feedback has reduced hallucinations by 37\% and improved user satisfaction and learning outcomes \cite{Jacobs_2024}.

\subsubsection{Cross-domain, comparative, multi-modal feedback integration}

LLM-Cure addresses large-scale user review analysis by combining automated feature extraction with competitive analysis \cite{assi2024llmcurellmbasedcompetitoruser}. Analyzing over one million reviews from 70 apps, it introduces a three-stage remediation process. The method achieves an 85\% F1-score in categorizing feature-specific complaints, outperforming prior approaches by 13\%. The system matches issues with top competitor solutions and generates actionable suggestions, 73\% of which have been adopted by developers. Additionally, 68\% of recommended features appeared in app updates within three months, demonstrating effective, scalable, and market-aware remediation.

Complementing this, USER-LLM integrates textual feedback with behavioral data—such as clickstreams and session durations—into unified embeddings \cite{ning2024userllmefficientllmcontextualization}. This approach improves personalization accuracy by 28\% over text-only models and enables LLMs to distinguish between transient frustrations and systemic issues needing intervention.

\subsubsection{Feedback driven alignment}

Effective feedback-driven learning requires addressing challenges in feedback collection, interpretation, and alignment. Subjective user preferences, inconsistent evaluation criteria, and underrepresentation in feedback datasets can introduce bias and reduce generalizability \cite{christiano2023deepreinforcementlearninghuman, stiennon2022learningsummarizehumanfeedback}. To address this, recent research expands feedback modalities to include natural language critiques, numerical ratings, and direct edits, increasing expressiveness but complicating integration \cite{ouyang2022traininglanguagemodelsfollow}.

To overcome limitations of traditional RLHF, alternative alignment strategies have emerged. Direct Preference Optimization (DPO) simplifies training by replacing reinforcement learning with a closed-form objective, reducing computational overhead \cite{rafailov2024directpreferenceoptimizationlanguage}. Safe RLHF introduces dual objectives—helpfulness and harmlessness—to promote aligned and safe outputs \cite{dai2023saferlhfsafereinforcement, ji2025saferlhfvsafereinforcement}. Reinforcement Learning from AI Feedback (RLAIF) further automates alignment by using AI-generated reward signals, enhancing scalability \cite{lee2024rlaifvsrlhfscaling}.

\subsection{Our contribution}

This paper introduces a dynamic framework for enhancing LLMs through real user feedback, enabling continuous system-level adaptation. Unlike static RLHF pipelines, our method classifies, aggregates, embeds, and retrieves feedback within the generation loop. Validated using the \textit{LLM-as-a-Judge} protocol \cite{azov2024selfimprovingcustomerreviewresponse}, the framework improves contextual relevance, factual accuracy, and alignment. By integrating diverse feedback types—such as textual corrections and behavioral signals—and employing efficient optimization, we support scalable, robust, and user-centered LLM solutions.

\section{Methods}

\subsection{FLARE Framework Overview}

We present FLARE (Feedback-driven LLM Adaptive RAG Enhancement), an iterative framework designed to improve RAG systems by leveraging structured user feedback. FLARE operates through two independent phases: offline preprocessing and online inference. These phases run in parallel with the primary RAG pipeline. We evaluate the framework using three datasets: the Fictional Persona Dialogs dataset, available on Hugging Face  \href{https://huggingface.co/datasets/TPelc/Fictional_Persona_Dialogs_Anonymized-benchmark}{\texttt{Fictional\_Persona\_Dialogs\_Anonymized}}; the Trivia dataset, accessible on Hugging Face  \href{https://huggingface.co/datasets/TPelc/Current_Trivia_Knowledge-benchmark}{\texttt{Current\_Trivia\_Knowledge-benchmark}} ; and the Wireless Technical Documentation dataset (proprietary). The effectiveness is assessed using the \textit{LLM-as-a-Judge} protocol.
\subsection{Feedback processing architecture}

\subsubsection{Offline feedback management pipeline}

The offline component of FLARE processes user feedback through a multi-stage pipeline that extracts actionable insights and converts them into structured knowledge representations:

\paragraph{Feedback collection and storage}
User-generated free-text feedback is continuously collected and stored in a MongoDB database. This feedback is processed in periodic batches to ensure scalability while maintaining a responsive user experience.

\paragraph{Contextual enrichment}
Raw feedback often lacks sufficient context for standalone interpretation. To address this, each feedback instance is enriched with relevant conversation history, converting fragmented inputs into self-contained, interpretable statements. For example:
\begin{itemize}
    \item Original interaction: Q: ``Who is the current president of the US?'' A: ``Joe Biden.'' F: ``Wrong, it is Donald Trump.''
    \item Enriched feedback: ``The current president of the US is Donald Trump as of 2025.''
\end{itemize}

\begin{figure}
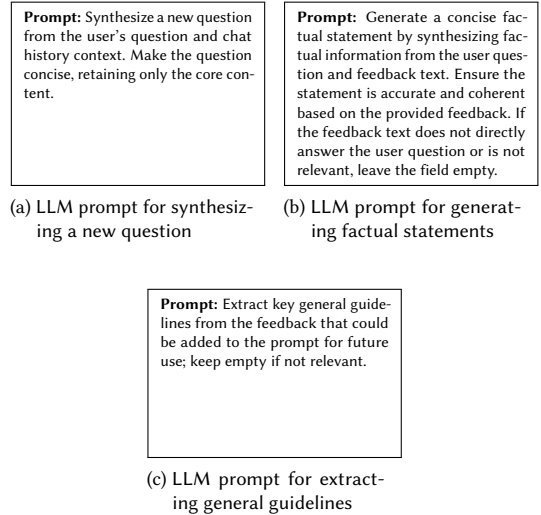

    \centering
    \subfloat[LLM prompt for synthesizing a new question]{%
        \begin{minipage}{0.4\linewidth}
            \scriptsize
            \centering
            \vspace{0.2cm}
            \fbox{
                \begin{minipage}[t][2.2cm][t]{\linewidth}
                    \textbf{Prompt:} Synthesize a new question from the user's question and chat history context. Make the question concise, retaining only the core content.
                \end{minipage}
            }
        \end{minipage}
        \label{fig:prompt_synthesize}
    }
    \hspace{0.05\linewidth}
    \subfloat[LLM prompt for generating factual statements]{%
        \begin{minipage}{0.4\linewidth}
            \scriptsize
            \centering
            \vspace{0.2cm}
            \fbox{
                \begin{minipage}[t][2.2cm][t]{\linewidth}
                    \textbf{Prompt:} Generate a concise factual statement by synthesizing factual information from the user question and feedback text. Ensure the statement is accurate and coherent based on the provided feedback. If the feedback text does not directly answer the user question or is not relevant, leave the field empty.
                \end{minipage}
            }
        \end{minipage}
        \label{fig:prompt_factual}
    }

    \subfloat[LLM prompt for extracting general guidelines]{%
        \begin{minipage}{0.4\linewidth}
            \scriptsize
            \centering
            \vspace{0.2cm}
            \fbox{
                \begin{minipage}[t][2cm][t]{\linewidth}
                    \textbf{Prompt:} Extract key general guidelines from the feedback that could be added to the prompt for future use; keep empty if not relevant.
                \end{minipage}
            }
        \end{minipage}
        \label{fig:prompt_guidelines}
    }
    \caption{System prompt instructions. The figure illustrates different types of prompts used in LLM systems for synthesizing questions, generating factual statements, and extracting guidelines.}
    \label{fig:prompts}
\end{figure}

\paragraph{Multi-category feedback classification}
The enriched feedback is automatically processed by LLM to generate two categories:

\begin{enumerate}
    \item \textbf{Factual data corrections}— Target inaccuracies, outdated information, or missing knowledge, prompting direct updates to the system’s knowledge base (Figure~\ref{fig:prompt_factual}).
    \item \textbf{Stylistic and structural guidelines}— Address clarity, coherence, and presentation quality without altering factual content (Figure~\ref{fig:prompt_guidelines}).
\end{enumerate}

\paragraph{Query synthesis and refinement}
Original user queries undergo context-aware refinement to remove extraneous elements while preserving core semantic meaning, as illustrated in Figure~\ref{fig:prompt_synthesize}. This process ensures feedback can be accurately linked to clean, standardized query representations.

%\begin{figure}[h]

\paragraph{Structured knowledge organization}
Processed outputs are stored as structured dictionaries, pairing synthesized queries with corresponding factual corrections and stylistic guidelines. This format enables efficient retrieval and application during inference.

\paragraph{Intelligent feedback aggregation}
GPT-4o aggregates feedback by identifying and consolidating redundant or conflicting entries. A relevance filter retains only actionable feedback that directly improves system performance, optimizing computational efficiency.

\paragraph{Vector embedding and indexing}
Synthesized queries are embedded using OpenAI’s text-embedding-ada-002 model and indexed in an Azure Vector Database, enabling fast similarity-based retrieval during inference.

\begin{figure*}
    \centering
    \includegraphics[width=\textwidth]{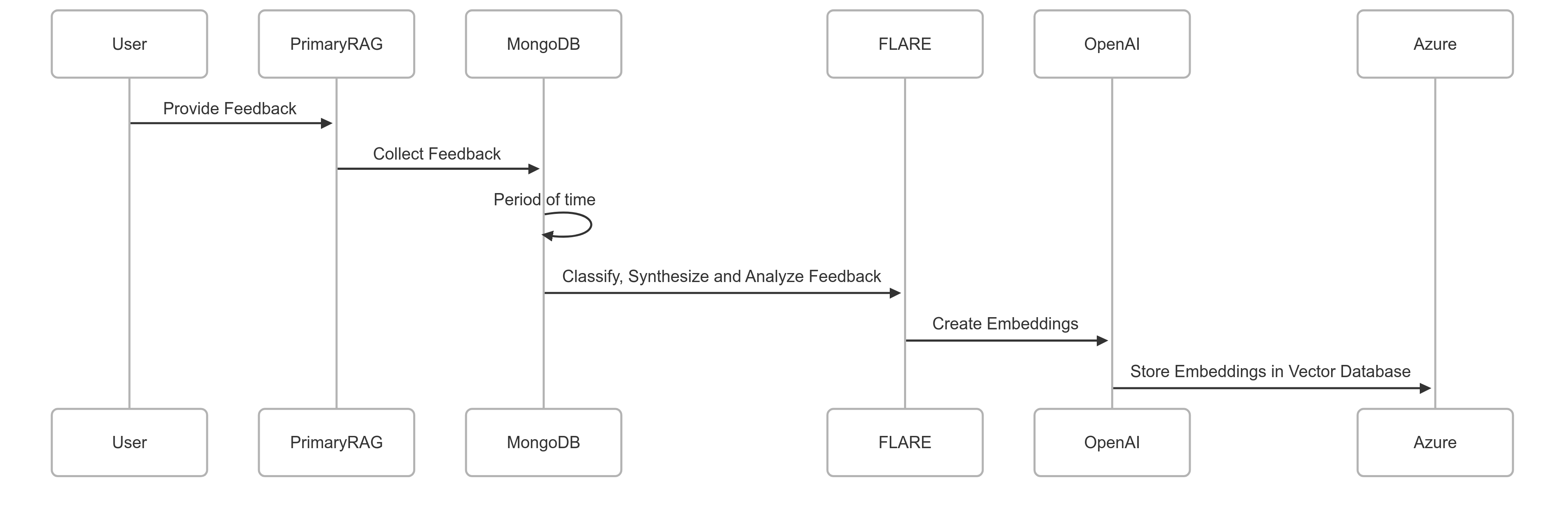} 
    \caption{FLARE offline processing flow}
    \label{fig:FLARE_offline}
\end{figure*}

\subsubsection{Online feedback integration workflow}

The online component integrates processed feedback into the active RAG pipeline in real time:

\paragraph{Dual retrieval architecture}
Each incoming query is processed in parallel by the primary RAG system and a dedicated Feedback-RAG module. This setup ensures comprehensive coverage without increasing latency.

\paragraph{Adaptive feedback integration}
Azure Hybrid Search identifies previously indexed queries that exhibit semantic similarity to the current user input. Queries exceeding a predefined relevance threshold are selected for integration. Factual corrections from these selected queries are injected directly into the primary RAG context, while stylistic guidelines are integrated as system-level instructions to refine output formatting and tone.

%\begin{figure*}[htbp]
\begin{figure*}
    \centering
    \includegraphics[width=0.66\textwidth]{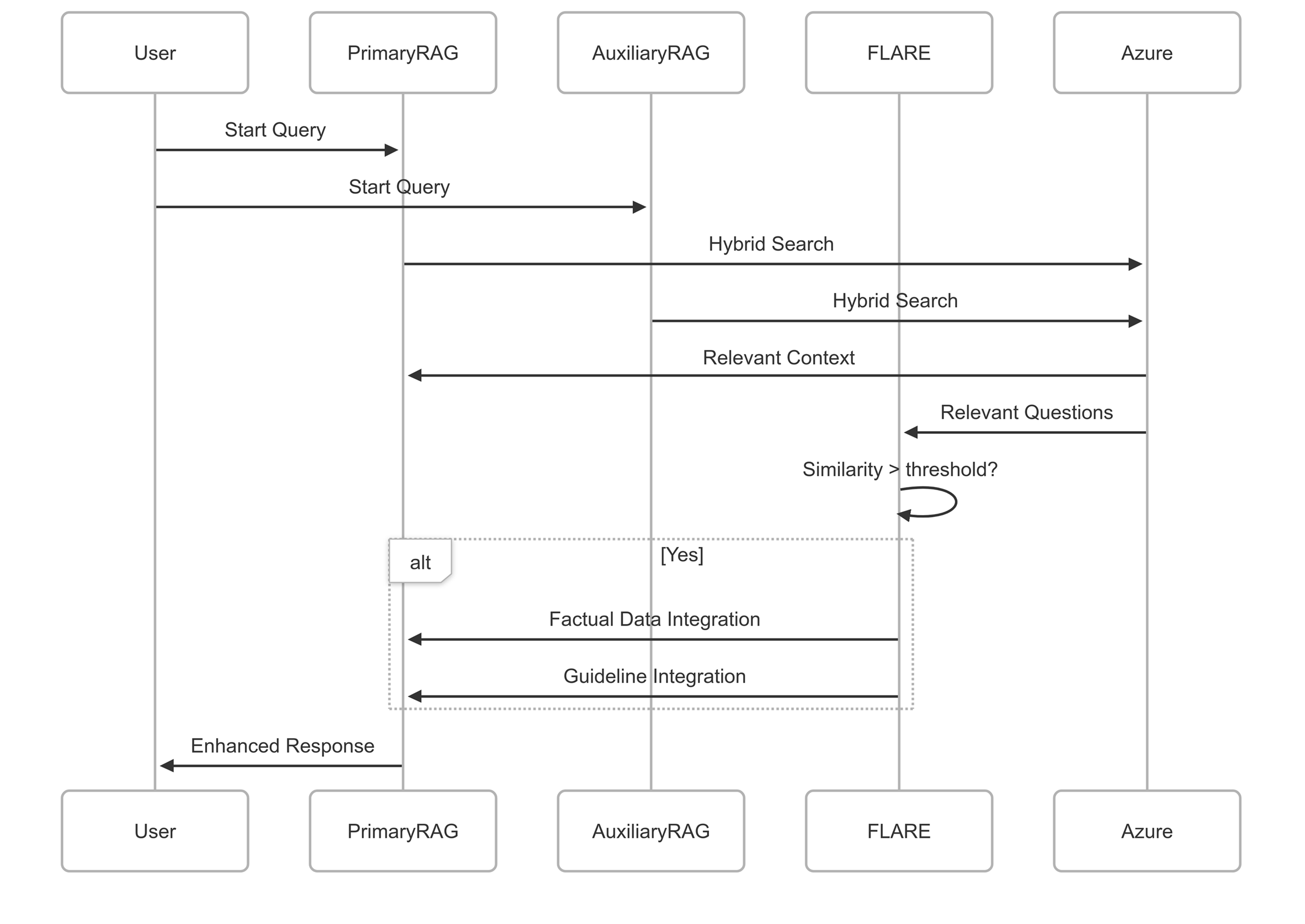} 
    \caption{FLARE inference flow}
    \label{fig:FLARE_inference}
\end{figure*}
%\end{figure*}

\begin{algorithm}[!t]
  \caption{Batch processing of user feedback (offline FLARE)}
  \KwIn{Batch $B = \{ f_{1}, \dots, f_{n} \}$ of raw user feedback; conversation log $\mathcal{C}$}
  \KwOut{Structured feedback store $\mathcal{S}$; vector index $\mathcal{V}$}

  \BlankLine
  \ForEach{$f_i \in B$}{
        $\tilde{f}_i \leftarrow \textsc{EnrichContext}(f_i, \mathcal{C})$\tcp*[l]{Add dialogue history}
        $(q_i, \kappa_i) \leftarrow \textsc{RefineQueryAndExtractCore}(\tilde{f}_i)$\;
        $c_i \leftarrow \textsc{ClassifyFeedback}(\kappa_i)$\tcp*[l]{factual vs. stylistic}
        $\mathcal{S} \leftarrow \mathcal{S} \cup \{ (q_i, c_i) \}$\;
  }
  $\mathcal{S} \leftarrow \textsc{AggregateAndResolveConflicts}(\mathcal{S})$\;
  $\mathcal{S} \leftarrow \textsc{FilterActionable}(\mathcal{S})$\tcp*[l]{drop noisy items}
  
  \BlankLine
  \ForEach{$(q, c) \in \mathcal{S}$}{
        $v \leftarrow \textsc{Embed}(q)$\;
        $\mathcal{V} \leftarrow \mathcal{V} \cup \{ (v, c) \}$\tcp*[l]{store in Azure Vector DB}
  }
  \Return{$(\mathcal{S}, \mathcal{V})$}
\end{algorithm}

\begin{algorithm}[!t]
  \caption{Adaptive feedback-aware inference (online FLARE)}
  \KwIn{User query $q^{\ast}$; primary RAG system $\mathcal{R}$; feedback store $\mathcal{S}$; vector index $\mathcal{V}$; similarity threshold $\tau$}
  \KwOut{Final answer $a^{\ast}$ with integrated factual and stylistic improvements}

  \BlankLine
  $v^{\ast} \leftarrow \textsc{Embed}(q^{\ast})$\;
  $\mathcal{F} \leftarrow \textsc{HybridSearch}(\mathcal{V}, v^{\ast}, \tau)$\tcp*[l]{retrieve nearest neighbours}
  
  \If{$\mathcal{F} \neq \emptyset$}{
        $(\Delta_\text{fact}, \Delta_\text{style}) \leftarrow \textsc{ExtractCorrections}(\mathcal{F}, \mathcal{S})$\;
        $\mathcal{R} \leftarrow \textsc{InjectFactualContext}(\mathcal{R}, \Delta_\text{fact})$\;
        $\mathcal{R} \leftarrow \textsc{InjectStylisticGuidelines}(\mathcal{R}, \Delta_\text{style})$\;
  }
  
  \BlankLine
  $a^{\ast} \leftarrow \mathcal{R}(q^{\ast})$\tcp*[l]{run augmented RAG}
  \textsc{LogInteraction}($q^{\ast}, a^{\ast}$)\tcp*[l]{for future offline batches}
  
  \Return{$a^{\ast}$}
\end{algorithm}

\subsection{Experimental design and datasets}

\subsubsection{Dataset selection and construction}

We evaluated FLARE on three datasets representing varying degrees of real-world complexity, from fully synthetic environments to real user interactions and feedback:

\paragraph{Fictional Persona Dialogs (FPD) dataset}
This dataset consists of 68 anonymized dialogues from fictional media sources, selected to avoid contamination from prior model knowledge. Question-answer pairs were generated contextually using GPT-4o, and the dataset is publicly available on Hugging Face for reproducibility.

Baseline evaluations without contextual information demonstrated low accuracy (\textasciitilde1 on a 1–5 scale), which improved dramatically to near-perfect performance following context integration. For FLARE evaluation, we removed 30\% of the context to measure improvements in a controlled setting. Full-context answers were used to generate training feedback, while semantically altered questions formed the test set—ensuring improvements reflected adaptive generalization, not memorization

\paragraph{Trivia dataset}
This dataset simulates realistic RAG deployment conditions involving ambiguous, time-sensitive queries and human feedback. It spans diverse domains—e.g., Olympic results, Nobel laureates, leadership changes—requiring up-to-date knowledge.

Question-answer pairs were generated with GPT-4o using recent web-sourced data; context was drawn from relevant Wikipedia articles to reflect standard RAG usage. The dataset includes 70 training and 30 testing question-answer pairs. Human-annotated feedback guided the training phase, while test question-answer pairs were derived from feedback-enhanced examples to evaluate generalization.

\paragraph{Wireless Technical Documentation dataset}
This domain-specific dataset reflects real-world RAG usage in technical support, requiring expert knowledge of wireless networking. Questions were sourced from actual user queries; answers were generated using retrievals from Intel documentation, specifications, and internal resources.
The dataset contains 80 question-answer pairs for training. Test queries were semantically modified versions of training examples to assess generalization and robustness in a high-precision, expert-driven domain.

\begin{table*}
  \caption{Summary of evaluation datasets}
  \label{tab:datasets}
  \begin{tabular}{lllll}
    \toprule
    \textbf{Dataset} & \textbf{Size (Train/Test)} & \textbf{Domain} & \textbf{Source Type} & \textbf{Purpose} \\
    \midrule
    Fictional Persona Dialogs & 68 (train/test split) & Open-domain dialogue & Synthetic & Controlled evaluation \\
    Trivia  & 70/30 & Current events & Semi-synthetic & Generalization \\
    Wireless Technical Documentation & 80 (train/test split) & Technical Q\&A & Real-world & Domain-specific validation\\ 
    \bottomrule
  \end{tabular}
\end{table*}

\subsection{Evaluation framework}

\subsubsection{\textit{LLM-as-a-Judge} assessment protocol}

We implemented a comprehensive evaluation framework using an \textit{LLM-as-a-Judge} approach to systematically assess response quality across multiple dimensions:

\paragraph{Quantitative metrics}
For list-based queries requiring enumeration or categorization, we used F1-score metrics to balance precision and recall, based on standard True Positive (TP), False Positive (FP), and False Negative (FN) counts. Representative examples include: ``List all products implementing Wi-Fi 6 technology'' and ``Which frame types can remain unprotected, and under what conditions are they dropped?''.

\paragraph{Qualitative assessment criteria}
Non-list queries underwent comprehensive evaluation across three critical dimensions:
\begin{enumerate}
    \item \textbf{Factual Accuracy}: Verification of content correctness against established ground truth
    \item \textbf{Contextual Relevance}: Assessment of response appropriateness to the specific query context
    \item \textbf{Completeness}: Evaluation of response comprehensiveness relative to predefined reference answers
\end{enumerate}

This multi-dimensional evaluation provides robust validation of FLARE's effectiveness and establishes a reproducible framework for future research in adaptive retrieval and feedback-driven generative AI.

\section{Results}

As described in the previous section, we constructed three benchmark datasets to gain confidence on the premise of our solution across both general and domain-specific knowledge settings, demonstrating FLARE’s versatility and effectiveness in diverse contexts..

We generated embedding representations using question-answer conversations and feedback history, and synthesized revised questions to form a test set. Below, we report our findings, showing significant gains in RAG system performance when augmented with the FLARE methodology.

\begin{table*}
  \caption{Example of how FLARE leverages user feedback: initial RAG response and score; injected knowledge and instruction; improved output with its correctness score.}
  \label{tab:flare-example}
\scriptsize
  \centering
  \rowcolors{2}{gray!10}{white} % alternate shading for readability
  \begin{tabular}{p{0.25\linewidth} p{0.68\linewidth}}
    \toprule
    \textbf{Section}                  & \textbf{Content} \\
    \midrule
    \textbf{User Question}            &  "Which acting award recipient at the 2025 Oscars both earned their very first nomination and became the first American of Dominican descent ever to win an acting Oscar?" \\
    \addlinespace
    \textbf{Reference Answer}         & "Zoe Saldaña, another first-time nominee, took home Best Supporting Actress for her performance in *Emilia Pérez*, making history as the first American of Dominican descent to win an acting Oscar." \\
    \addlinespace
    \textbf{Initial RAG Response}     & "The acting award recipient at the 2025 Oscars who both earned their very first nomination and became the first American of Dominican descent ever to win an acting Oscar is Mikey Madison. She won the award for Actress in a Leading Role for her performance in \textit{Anora}"\\
    \addlinespace
    \textbf{Suggested Knowledge}      & "At the 2025 Oscars, several first-time nominees won major awards, including Mikey Madison, Kieran Culkin, and Zoe Saldaña, who made history as the first American of Dominican descent to win an acting Oscar. Sean Baker also tied Walt Disney’s record for most wins in a single night, with additional debut wins in directing and technical categories." \\
    \addlinespace
    \textbf{Suggested Takeaway}       & "Include detailed and historically significant information from the feedback when summarizing key winners, such as notable upsets, demographic records, and additional category recognitions." \\
    \addlinespace
    \rowcolor{ImprovedRow}
    \textbf{RAG Response with FLARE}  & "Zoe Saldaña earned her very first nomination and became the first American of Dominican descent ever to win an acting Oscar, winning in the category of Actress in a Supporting Role for her performance in \textit{Emilia Pérez}." \newline\textbf{Score:} 5 (correct host, country, and dates) \\
    \bottomrule
  \end{tabular}
  \end{table*}

\subsection{General knowledge: FPD benchmark}
\begin{figure*}[htbp]
  \centering
  \includegraphics[width=\textwidth]{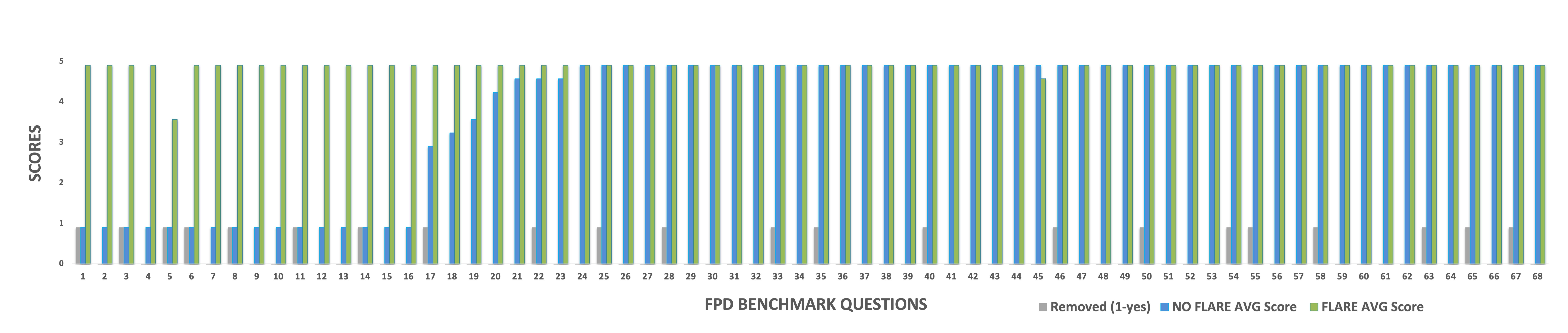}
  \caption{Performance on the 68-question FPD benchmark, with and without FLARE.  
           FLARE attains the maximum score of~5 on all but two items, which receive
           scores of~3.66 and~4.67, respectively (scores are out of~5).}
  \label{fig:FPDComparison}
  \vspace{-5pt} % Optional: tighten vertical space after the figure
\end{figure*}

This benchmark was primarily used as a sanity check to verify the correctness of our implementation and ensure adherence to the intended FLARE flow. 
As can be seen in the results depicted in Figure \ref{fig:FPDComparison} and Table \ref{tab:FPD} , we  demonstrate a marked improvement in performance  with the use of \textit{FLARE}, achieving a maximum score of 5 on nearly all items.

\begin{table}[h]
\centering
\caption{FPD benchmark results summary}
\label{tab:FPD}
\begin{tabular}{lccc}
\toprule
            & \multicolumn{2}{c}{Score} & T-Test (p) \\ 
\cmidrule(lr){2-3}
            & No FLARE & FLARE &  \\ 
\midrule
AVG         & 4.0      & 5.0    & --      \\
STD         & 1.7      & 0.2    & \(4 \times 10^{-6}\) \\
\bottomrule
\end{tabular}
\end{table}

\subsection {General knowledge: Trivia benchmark}

Additionally, we constructed a benchmark consisting of Trivia questions centered on events from 2024 and 2025—beyond the knowledge cut-off of the LLM under evaluation (GPT-4o). To ensure the RAG system had access to relevant, up-to-date information, we supplemented its retrieval corpus with curated Wiki and web pages. Human evaluators were then asked to provide textual feedback on the system’s responses. Using the questions, RAG answers, and feedback history, we generated embedding representations and created a revised set of test questions within the same domain.  
\begin{figure}[htbp] % 'htbp' for positioning
  \centering
  \includegraphics[width=\linewidth]{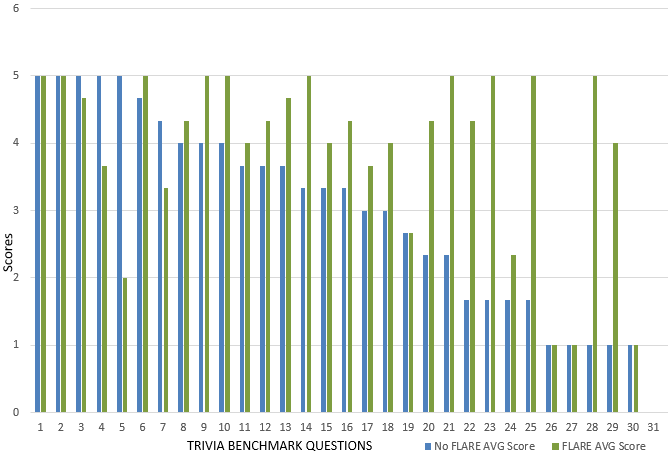}
  \caption{Results with and without FLARE for 30  Trivia benchmark questions: where FLARE average score is 3.91 versus NO FLARE 3.06, respectively (out of score 5)}
 \label{fig:TriviaComparison}
\end{figure}
%\vspace{-5pt} % Adjust the value as needed

The integration of FLARE resulted in a substantial improvement in both the accuracy and relevance of responses. As shown in Figure~\ref{fig:TriviaComparison}, FLARE achieved an average score of 3.91 compared to 3.06 for the baseline RAG system, based on a test set of 30 newly generated questions from the same domain as the training data. Average and best score were computed across three different responses per question. Unlike the baseline RAG system—which struggled to generate accurate answers even with access to up-to-date Wiki and web content—FLARE effectively leveraged user feedback to adapt to new information beyond its initial feedback training data.
Table~\ref{tab:Triviatestset} presents a curated sample of test questions, underscoring the LLM's efficacy in evaluating the quality of RAG system responses. The table also highlights the substantial enhancement in answer quality when utilizing FLARE compared to the absence of FLARE.

\begin{enumerate}[leftmargin=*] % Adjusts list indentation
\item \textbf{Feedback utilization:} FLARE’s dynamic feedback retrieval mechanism plays a central role across benchmarks, enabling continuous adaptation based on user-provided feedback. This iterative loop yields more contextually appropriate and nuanced responses.
\item \textbf{ \textit{LLM-as-a-Judge} evaluation:} Implementing this evaluation strategy offers a robust, scalable framework for assessing FLARE's improvements. It demonstrates the system's alignment with human judgment and preferences, further validating the efficacy of the feedback-driven approach.
\item \textbf{Targeted context via feedback:} Sample question-answer pairs reveal that the baseline RAG system often fails to produce satisfactory responses, even when relevant data is available. In contrast, FLARE effectively utilizes context extracted from previous user feedback to provide satisfactory answers.
\end{enumerate}

\begin{table*}[htbp]
        \caption{\textbf{Sample Trivia test set evaluation via LLM:} Each entry presents the user query, the RAG system's response without FLARE, followed by feedback and the response with FLARE.}
        \centering
    \scriptsize % Reduce font size for the table
    \begin{tabular}{|p{1\linewidth}|} % Adjusted column width
        \hline
        \textbf{RAG Answer with and without FLARE} \\
        \hline
        \textit{\textcolor{blue}{Who won the Best Actor award at the 2025 Oscars, and for which film performance was the accolade given?}} \\
        \textbf{No FLARE:} Adrien Brody won the Best Actor award at the 2025 Oscars for his performance in "The Brutalist." \\
        \textbf{LLM evaluation:} The response is accurate regarding Adrien Brody's win for "The Brutalist." \\
        \textbf{FLARE:} Adrien Brody won the Best Actor award for his role as László Tóth in "The Brutalist." This marks his second Oscar win, the first being for "The Pianist" in 2003. \\
        \hline
        \textit{\textcolor{blue}{Which company or individual won the Best Innovation Award at CES 2025, and what innovation did they showcase?}} \\
        \textbf{No FLARE:} Anker Solix won the Best Innovation Award at CES 2025 for their solar-powered cloak. This jacket blends fashion with technology, allowing device charging on the go. \\
        \textbf{LLM evaluation:} The response incorrectly mentions Anker Solix, while Suprema AI won the award for their Q-Vision Pro. \\
        \textbf{FLARE:} Suprema AI won the Best Innovation Award for their Q-Vision Pro, an AI module for predicting and preventing financial crimes at ATMs. It features AI-powered facial recognition and behavior analysis. \\
        \hline
        \textit{\textcolor{blue}{What is the official name of Intel's PC AI platform launched in late 2024-2025, and what are its main features?}} \\
        \textbf{No FLARE:} Gemini is Intel's PC AI platform, featuring AI-powered performance optimization and enhanced security. It integrates seamlessly with other devices. \\
        \textbf{LLM evaluation:} The response incorrectly names the platform "Gemini," while it is officially the Intel Core Ultra processor family. \\
        \textbf{FLARE:} Intel's PC AI platform, the Intel Core Ultra processor family, integrates CPUs, GPUs, and NPUs to enhance productivity and security. The Lunar Lake series marks a significant step in embedding AI capabilities into hardware. \\
        \hline
        \textit{\textcolor{blue}{Did Intel win a Best Product Award at MWC 2025, and what was the specific category of the award?}} \\
        \textbf{No FLARE:} The context does not mention Intel winning a Best Product Award at MWC 2025. Various gadgets were showcased, but Intel's award status is unconfirmed. \\
        \textbf{LLM evaluation:} The response incorrectly states Intel did not win an award, while they won the "Best AI Debut Award" for their Intel AI Assistant Builder Tool. \\
        \textbf{FLARE:} Intel won the "Best AI Debut Award" for their Intel AI Assistant Builder, a tool for creating customized AI assistants. This highlights Intel's innovation in making AI technology more accessible. \\
        \hline
    \end{tabular}
    \label{tab:Triviatestset}
\end{table*}
\vspace{-10pt} % Adjust the value as needed

\subsection {Custom domain benchmark}
In the custom Wireless domain, we built a benchmark from a year’s worth of real-world customer feedback on our internal large-scale RAG platform. It reflects practical challenges and evolving needs in a dynamic environment.

For this benchmark, FLARE effectively integrated user feedback to enhance the RAG system's responses. We embedded 210 textual user feedback entries collected over the year. These entries were provided by users in response to a dislike tag and were integrated into a RAG system built on division-specific data.

We created a test set of 80 questions, derived from the original user queries but phrased differently. We assessed FLARE's ability to utilize feedback for more accurate answers in two key areas:
\begin{enumerate}[leftmargin=*] % Adjusts list indentation
\item \textbf{Ability to extract relevant feedback:} we  initially tested FLARE’s ability to retrieve relevant feedback for queries related to past user input. In our custom test set, FLARE successfully extracted relevant feedback in 95\% of cases (76 out of 80 queries). The four failures occurred with queries concerning specific JIRA entries, where feedback linked to the same entries was not always retrieved.
\item \textbf{Ability to enhance answers using feedback:} we further evaluated FLARE's ability to generate meaningful answers by leveraging feedback, automating this process using an \textit{LLM-as-a-Judge}. Results showed that although relevant feedback was retrieved as context in nearly all cases (95\%), FLARE incorporated that feedback into its answers only 61\% of the time.
\end{enumerate}

Our results clearly demonstrate the substantial improvements made possible by incorporating user feedback into the continuous self-improvement cycle of RAG systems. Nevertheless, as illustrated in Table~\ref{tab:Customtestset}, there remain opportunities to further refine FLARE’s performance.

\begin{table*}[htbp]
   \caption{\textbf{Insights for enhancing FLARE:} this table presents three examples from a custom domain benchmark, highlighting challenges and potential improvements for FLARE (user feedback-driven RAG). Generic names are used to protect intellectual property. The table also includes insights for further development, identifying areas where FLARE can be improved.}
      \centering
    \scriptsize % Reduce font size for the table
    \begin{tabular}{|p{1\linewidth}|} % Adjusted column width
        \hline
        \textbf{Insights on further improvements of FLARE answers (based on detailed analysis of test results from the custom test set)} \\
        \hline
        \textit{\textcolor{blue}{In which JIRA entry is Feature X referenced?}} \\
        \textbf{Original User Feedback:} Feature X appears in JIRA Entry-61987 but was not referenced for some reason. \\
        \textbf{FLARE:} Feature X is referenced in JIRA Entry-61987. \\
        \textbf{Insight:} \textit{Retrieval Reiteration} - The FLARE answer could be further improved by providing details on the respective JIRA entry. This can be achieved by reiterating the RAG flow once context for the specific user query is obtained from multiple sources, such as feedback and vector store embedding of custom domain data. \\
        \hline
        \textit{\textcolor{blue}{Which specific models does OEM Y use? Please refer to models (a, b, and c) in your answer.}} \\
        \textbf{Original User Feedback:} Only models a and b are used by OEM Y. The c model is used by OEM X. \\
        \textbf{FLARE:} OEM Y uses several versions of models a, b, and c. \textit{\textless Rest of the answer \textgreater} \\
        \textbf{Insight:} \textit{Feedback Prioritization} - The fact that OEM Y does not use model c was overlooked. This indicates that the information in the feedback needs to be prioritized. \\
        \hline
        \textit{\textcolor{blue}{Please provide details of architectural feature Z.}} \\
        \textbf{Original User Feedback:} Should at least provide some pointers to the right JIRA entry for this architectural feature. \\
        \textbf{FLARE:} The details of the architectural feature are... \textit{\textless Details of the architectural feature are provided without pointers to specific JIRA entries \textgreater} \\
        \textbf{Insight:} \textit{Feedback Generalization} - The FLARE answer could be improved through generalized learning from all past feedbacks, such as always including JIRA entry information when providing details on architectural features. \\
        \hline
    \end{tabular}
    \label{tab:Customtestset}
\end{table*}
\vspace{-3pt} % Adjust the value as needed

\section{Conclusions and future work}
We introduced FLARE, a feedback-aware extension to RAG.  
A straightforward pipeline—comprising feedback classification, aggregation, embedding, and retrieval at a similarity threshold of 0.85—significantly improves average answer quality, as demonstrated across three diverse contextual benchmarks.  
The use of \textit{LLM-as-a-Judge} aligns automated scoring with human preference while offering concise rationales.

During evaluation, two notable limitations emerged:
\textbf{First}, although the retrieval module successfully located relevant feedback for nearly every custom-domain query, the generation module incorporated this feedback in only 61\% of the answers.  
This \emph{retrieval–utilization gap} suggests that our current context injection method—template-based string concatenation—can marginalize important details.
We currently lack a principled mechanism to prioritize feedback when model attention is constrained, or when snippets overlap or conflict.

\textbf{Second}, feedback selection currently depends on a fixed cosine similarity threshold and treats all retrieved comments as carrying equal weight.  This uniform policy ignores cues that strongly influence a comment’s utility—such as author expertise, recency, consistency with other feedback, and historical impact on answer quality.  As a result, the system may surface outdated or low-quality guidance while excluding more recent, reliable corrections that fall just below the fixed cut-off.
In future work, we plan to develop adaptive ranking mechanisms that weight feedback based on relevance, freshness, and author credibility. We also aim to incorporate an iterative decoding loop that re-retrieves evidence when the current context is sparse or contradictory. Additionally, we intend to broaden the feedback schema to support hierarchical labels and multimodal inputs, such as screenshots, voice notes, and click traces. Finally, we will build a low-latency streaming pipeline capable of ingesting, evaluating, and applying feedback in real time, blending automated scoring with targeted expert review.

Together, these extensions aim to advance FLARE toward becoming a fully autonomous, continuously self-improving RAG framework.

%% The declaration on generative AI comes in effect
%% https://ceur-ws.org/GenAI/Policy.html
\section*{Declaration on Generative AI}
 During the preparation of this work, the author(s) used GPT-4o spell check and revise grammar, after which the author(s) reviewed and edited the content as needed and take(s) full responsibility for the publication’s content. 

%%
%% Define the bibliography file to be used
\bibliography{paper}

\end{document}